 \documentclass[final,3p,times]{elsarticle}

\usepackage{amssymb}

\usepackage{lineno}

\journal{Journal Name}

\newtheorem{thm}{Theorem}
\newtheorem{lem}{Lemma}
\newdefinition{rmk}{Remark}

\newproof{pf}{Proof}
\newproof{pot}{Proof of Theorem \ref{thm2}}

\begin{document}

\begin{frontmatter}

\title{ Overlap Coefficients Based on Kullback-Leibler Divergence: Exponential Populations Case}

\author[mysecondaryaddress]{Hamza Dhaker\corref{mycorrespondingauthor}}
\cortext[mycorrespondingauthor]{Corresponding author}
\ead{hamzaould.yahya@ucad.edu.sn}
\author[mysecondaryaddress,mymainaddress]{Papa Ngom}
\author[mytherdaryaddress]{Malick Mbodj}
\address[mysecondaryaddress]{LMDAN,Universit\'{e} Cheikh Anta Diop, Dakar, Senegal}
\address[mymainaddress]{LMA,Universit\'{e} Cheikh Anta Diop, Dakar, Senegal}
\address[mytherdaryaddress]{Bowie State University, Maryland, USA}

\begin{abstract}
This article is devoted to the study of overlap measures of densities of two exponential populations. Various Overlapping Coefficients, namely: Matusita's measure $\rho$, Morisita's measure $\lambda$ and Weitzman's measure $\Delta$. A new overlap measure $\Lambda$ based on Kullback-Leibler measure is proposed. The invariance  property and a method of statistical inference of these coefficients also are  presented. Taylor series  approximation are used to construct confidence intervals for the overlap measures. The bias and mean square error properties of the estimators are studied through a simulation study.
\end{abstract}

\end{frontmatter}

\linenumbers

\section{Introduction}
\label{Intro}
The similarity between two densities can be considered as the commonality shared by both populations. Generally it is measured on the scale of $0$ to $1$. Values of measure close to $0$ corresponding to the distributions having supports with no intersection and $1$ to the perfect matching of the two distributions. Scientists from different disciplines propose different measures of similarity serving different purposes.
\\
By using delta method Smith \cite{17} derived formulas for estimating the mean and  the variance of the discrete version of Weizman's measure (also known as the overlap coefficient).
Mishra et al. \cite{11} gave the small and large sample properties of the sampling distributions for a function of this overlap measure estimator, under the assumption of homogeneity of variances for the case of two normal distributions. Mulekar and Mishra \cite{13} simulated the sampling distribution  of  estimators  of  the  overlap measures when the two densities correspond to the normal case with equal means and obtained the  approximate  expressions  for the bias and  variance of their estimators.
\\
Smith \cite{17} derived approximate formulas using the delta method for estimating the mean
and variance of the discrete version of one such measure known as Weitzman's
measure $\Delta$(Weitzman \cite{18}) (also known as the overlap coefficient). Mishra  et  al.  \cite{11}  gave  some properties  of  the  sampling  distributions  for  a function  of  the estimator,  under  the assumption of homogeneity of variances for the case  of  two  normal  distributions. Recently, several authors including Bradley and Piantadosi \cite{4}, Inman and Bradley \cite{7}, Clemons \cite{5}, Reiser and Faraggi \cite{16}, Clemons and Bradley \cite{6}, Mulekar and Mishra \cite{14}, Al-Saidy, et al. \cite{1},  Al-Saleh and Samawi \cite{2}, and Samawi and Al-Saleh \cite{016}  considered this measure.  
\\
 Dixon \cite{06} described the use of bootstrap and jackknife techniques for the Gini coefficient of size hierarchy, a commonly used measure of similarity between income distributions of two ethnic, gender, or geographical groups, and the Jaccard index of community similarity. AL-Saidy et al. \cite{1}  consider the problem of drawing inference about  the  three  overlap  measures under the Weibul distribution  function  with equal shape parameter. Wei Ning et al \cite{014} have compared  mixtures of generalized lambda distributions (GLDs) with normal mixtures by using KullbackLeibler $(KL)$ distance and overlapping coefficient $(\delta )$ .
\\
The main objective of this paper is to propose a new $OVL$ based on the Kulback-Leibler divergence \cite{8}  for two Exponential distributions, i.e. from a measure of divergence or dissimilarity, we construct a measure of similarity noted $\Lambda$ defined in (\ref{kll}). We provide its maximum likelihood estimator.
\\
The coefficients and their properties are given in section  2. The expressions  for  approximate  bias  and 
variance  of $OVL$  are  included  in  section  3.  A  method  for  making  statistical inferences about the
 $OVLs$ is also discussed in this section. The  results of simulation study  are described  in section 4, along 
 with an example demonstrating  the  usefulness  of $OVLs$.  Finally, the conclusion and perspective is presented  in Section 5.
 
\section{Overlap Coefficients}

We consider four different similarity measures (the overlap coefficients ($OVL$)): Matusita's measure $\rho$, Morisita's measure $\lambda$, Weitzman's  measure $\Delta$ and the measure based Kullback-Leibler divergence $\Lambda$. 
The overlap measure ($OVL$) is defined as the area of intersection of the graphs of two probability density functions. It measures the similarity, which is the agreement or the closeness of the two probability distributions.
\\
Let $F_{1}(x)$ and $F_{2}(x)$ be two distribution functions with the corresponding density functions with respect to the Lebesgue measure. Four commonly used measures that describe the closeness between $F_{1}(x)$ and $F_{2}(x)$ are described below;

\begin{itemize}

\item Weitzman's Measure \cite{18}  The overlapping coefficient $\Delta$ is the area under two functions simultaneously, defined as,
$$ \Delta = \int \min \left[ f_{1}(x),f_{2}(x)\right]dx. $$

\item Matusita's Measure \cite{10} second measure studied here is known as the Matusita's measure, $\rho$, which is defined as,

$$ \rho= \int \sqrt{f_{1}(x)f_{2}(x)}dx $$

This measure is based on the distance between two functions (Matusita \cite{10}). Matusita actually developed a discrete version of $\rho$, which is also known as the \textbf{Freeman-Tukey measure (FT)}. This
measure is related to the Hellinger distance (Rao \cite{15} and Beran \cite{3}). 

\item Morisita's Measure \cite{12} Morisita proposed an index of similarity between communities. Consider an ecological study involving two populations from each of which a random sample is taken, defined as,

$$ \lambda =\frac{2\int f_{1}(x)f_{2}(x)dx}{\int [f_{1}(x)]^{2}dx+\int [f_{2}(x)]^{2}dx} $$

\item Kullback-Leibler \cite{8} : The Kullback-Leibler divergence was originally introduced by Solomon Kullback and Richard Leibler in 1951 as the directed divergence between two distributions. It is discussed in Kullback's historic text, Information Theory and Statistics. \\
the overlap coefficient $\Lambda$ is the complement of Kullback-Leibler

\begin{eqnarray}
\label{kll}
 \Lambda = \frac{1}{1+KL(f_{1}\| f_{2} )}
 \end{eqnarray}
with $KL(f_{1}\| f_{2})= \int (f_{1}-f_{2})\log \left(\frac{f_{1}}{f_{2}} \right)dx  $
\end{itemize}

\subsection{Overlap measures (OVL) for Exponential Distribution }

The simplest and most commonly used distribution in survival and reliability analysis is the one-parameter exponential distribution. Suppose $f_{i}(x; \theta_{i})$ indicate
two exponential populations with respective hazard rates $\theta_{i}> 0(i=1,2)$, that is
$$ f_{i}(x; \theta_{i})=\theta_{i}\exp(\theta_{i}x), \quad for \quad x \in (0, \infty). $$

\begin{figure}
\begin{center}
\label{exponentiel}
\includegraphics[scale=0.7]{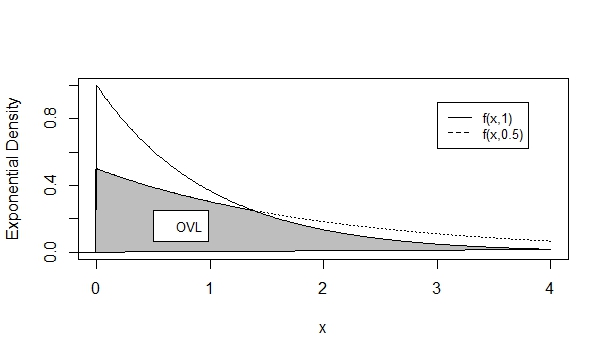} 
\end{center}
\caption{The overlap of two exponential densities.}
\end{figure}

The Overlapping Coefficients is shown graphically in Figure \ref{exponentiel}.
\\
Let $R = \frac{\theta_{1}}{\theta_{2}}$, the ratio of hazard rates, then these measures can be shown to be functions of $R$ as follows
 $$ \Delta = 1- \vert 1-\frac{1}{R} \vert R^{\frac{1}{1-R}} \qquad R\neq 1 $$

$$ \rho = \frac{2\sqrt{R}}{1+R} $$

$$ \lambda = \frac{4R}{(1+R)^{2}} $$
and
$$ \Lambda= \frac{R}{R^{2}-R+1} \qquad $$

\begin{lem}
For  OVLs  defined  earlier,
\begin{itemize}
\item[a)] $ 0\leq OVL \leq  1$ for all $R \geq 0$
\item[b)] $OVL=1$ iff $R=1$
\item[c)] $OVL=0$ iff $R=0$ or $R=\infty$
\end{itemize}
ll  four  OVLs  possess  properties  of  reciprocity,  invariance,  and  piecewise  monotonicity

\begin{itemize}
\item[a)] $OVL(R)= OVL(1/R) $
\item[b)] $OVLs$  are  monotonically  increasing  in $R$ for $0\leq R\leq 1$ and  decreasing  in $R>1$  
\end{itemize}
\end{lem}

\begin{figure}
\begin{center}
\label{exponentiel}
\includegraphics[scale=0.7]{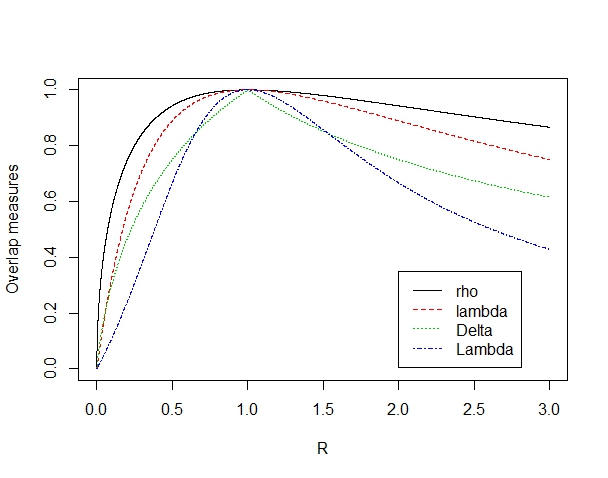} 
\end{center}
\caption{Measures of similarity as functions of R for exponential populations.}
\end{figure}

\section{Bias and Variance of Estimates}
\label{section2}

As  noted  earlier,  the  overlap  coefficients  are  functions  of  the  ratio.  Most  commonly, in  the  estimation  of  ratios,  estimators  that  are  convenient  and  easy  to  understand  are  found to  be  biased.  As  noted  by  Lu,  et  al.  (1989),  the  OVLs  in  this  study  are  no  exception  to  it. 
The  amount  of  bias  is  $B(OVL)=\mathbb{E}(OVL)-OVL$.  To  examine  the  effects  of  bias,  approximate  expressions  for  the  mean  and  the  variance  of  estimates  are  obtained.
\\
suppose that $(X_{ij}; j=1,...,n_{i}; i=1,2)$ denote independent observation from two independent random samples draw from $f_{1}(x)$ and $f_{2}(x)$ respectively, where
$$ f_{1}(x)=\frac{1}{\theta_{1}}\exp\lbrace -\frac{x}{\theta_{1}} \rbrace \qquad x >0 $$
 
and

$$ f_{2}(x)=\frac{1}{\theta_{2}}\exp\lbrace -\frac{x}{\theta_{2}} \rbrace \qquad x >0 $$

The maximum likelihood estimators (MLEs) based on the two samples are given by:

\begin{itemize}
\item[1)] From the first sample: 

$$ \widehat{\theta}_{1}= \overline{X}_{1}=\frac{\sum_{i=1}^{n_{1}}X_{1i}}{n_{1}} $$
\item[2)] From the second sample: 

$$ \widehat{\theta}_{2}= \overline{X}_{2}=\frac{\sum_{i=1}^{n_{2}}X_{2i}}{n_{2}} $$
\end{itemize}

Note that, it is easy to show that 

$$ \widehat{\theta}_{1}\sim G(n_{1}, \frac{\theta_{1}}{n_{1}}) \qquad \widehat{\theta}_{2}\sim G(n_{2}, \frac{\theta_{2}}{n_{2}}) $$
where 
$G(.,.)$  stands  for  the  gamma distribution function. Hence,  the  variances of those MLE's are respectively  
$ Var(\widehat{\theta}_{1})=\frac{\theta_{1}^{2}}{n_{1}}$ and $Var(\widehat{\theta}_{2})=\frac{\theta_{2}^{2}}{n_{2}} $ Then we may define an estimate of $R$ is $ \widehat{R}=\frac{\widehat{\theta}_{1}}{\widehat{\theta}_{2}} $.

Therefore, using the relationship between  Gamma distribution and Chi-square distribution and the fact that the two samples are independent, it is easy to show that $\frac{\theta_{1}}{\theta_{2}}\widehat{R}$ has $F$-distribution (i.e, $F(2n_{1},2n_{2})$).
Hence, the variance of $\widehat{R}$ is $ Var(\widehat{R})=R^{2}\frac{n_{2}^{2}(n_{1}+n_{2}-1)}{n_{1}(n_{2}-1)(n_{2}-2)} $ Also,  an unbiased  estimate  of  $R$  is  given by $ \widehat{R}^{*}= \frac{\widehat{\theta}_{1}}{\widehat{\theta}_{2}}\frac{(n_{2}-1)}{n_{2}}=\frac{(n_{2}-1)}{n_{2}}\widehat{R}$
with 
\begin{eqnarray}
\label{varR}
Var(\widehat{R}^{*})= R^{2}\frac{(n_{1}+n_{2}-1)}{n_{1}(n_{2}-2)} 
\end{eqnarray}.
 Clearly, $\widehat{R}^{*}$ has less variance than  $\widehat{R}$.

$$ \widehat{\Delta} = 1- \vert 1-\frac{1}{\widehat{R}^{*}} \vert (\widehat{R}^{*})^{\frac{1}{1-\widehat{R}^{*}}} \qquad  $$

$$ \widehat{\rho} = \frac{2\sqrt{\widehat{R}^{*}}}{1+\widehat{R}^{*}} $$

$$ \widehat{\lambda} = \frac{4\widehat{R}^{*}}{(1+\widehat{R}^{*})^{2}} $$
and
$$ \widehat{\Lambda}=\frac{\widehat{R}}{\widehat{R}^{2}-\widehat{R}+1} \qquad $$

\begin{thm}
Suppose $\widehat{\Delta}$, $\widehat{\rho}$, $ \widehat{\lambda}$ and $ \widehat{\Lambda}$ are the estimates of $\Delta$, $\rho$, $\lambda$ and $\Lambda$ respectively, obtained replacing $R$ by $\widehat{R}^{*}$.  the  approximate  sampling variance of the $OVL$ measures can be obtained as follows:

\begin{eqnarray}
Var(\widehat{\Delta}) = \frac{(n_{1}+n_{2}-1)(R)^{\frac{2}{1-R}}(\log R )^{2}}{n_{1}(n_{2}-2)(1-R)^{2}}
\end{eqnarray}

\begin{eqnarray}
Var(\widehat{\rho}) = \frac{R(1-R)^{2}(n_{1}+n_{2}-1)}{n_{1}(n_{2}-2)(1+R)^{4}}
\end{eqnarray}

\begin{eqnarray}
Var(\widehat{\lambda})=\frac{16R^{2}(1-R)^{2}(n_{1}+n_{2}-1)}{n_{1}(n_{2}-2)(1+R)^{6}}
\end{eqnarray}

\begin{eqnarray}
Var(\widehat{\Lambda})=\frac{(n_{1}+n_{2}-1)}{n_{1}(n_{2}-2)}\frac{R^{2}(1-R^{2})^{2}}{(R^{2}-R+1)^{4}}
\end{eqnarray}

\end{thm}

\begin{pf}
Since  each  of  the  $OVL$  is  a  function  of  $R$,  the  expressions  are  obtained  using the first  order  Taylor  series  expansion  about  $R$  and  the  $Var(\widehat{R}^{*})$ given in equation (\ref{varR}).
\end{pf}
\par

\begin{thm}
the  approximate  sampling bias of the$OVL$ measures can be obtained as follows: 
\begin{eqnarray}
Bias(\widehat{\Delta})= \displaystyle\left\{
    \begin{array}{l}
      \frac{(n_{1}+n_{2}-1)R^{2}}{n_{1}(n_{2}-2)}\frac{R^{\frac{2R-1}{1-R}}[R(2R-\log (R)-2 )\log (R) -(R-1)^{2} ]}{(R-1)^{3}} \quad if \quad R>1 \\ \\
    \frac{(n_{1}+n_{2}-1)R^{2}}{n_{1}(n_{2}-2)}\frac{R^{\frac{2R-1}{1-R}}[R(2R-\log (R)-2 )\log (R) -(R-1)^{2} ]}{(1-R)^{3}} \quad if \quad R<1 
 \end{array}
\right. 
\end{eqnarray}

\begin{eqnarray}
Bias(\widehat{\rho}^{*})=\frac{(n_{1}+n_{2}-1)\sqrt{R}}{n_{1}(n_{2}-2)}\frac{3R(R-2)-1}{2(R+1)^{3}}
\end{eqnarray}

\begin{eqnarray}
Bias(\widehat{\lambda}^{*})= \frac{n_{1}+n_{2}-1}{n_{1}(n_{2}-2)}\frac{8R^{2}(R-2)}{(R+1)^{4}}
\end{eqnarray}

\begin{eqnarray}
Bias(\widehat{\Lambda})=-\frac{n_{1}+n_{2}-1}{n_{1}(n_{2}-2)}\frac{R^{2}(2R^{3}-6R+2)}{(R^{2}-R+1)^{3}}
\end{eqnarray}
\end{thm}

\begin{pf}
Using the second order Taylor series expansion the desired results are obtained.
\end{pf}

\begin{rmk}
Reasonable  estimates  for  the  above  variances and the biases can be obtained by substituting $R$ by its consistency estimator $\widehat{R}^{*}$ in the above formulas.  
\end{rmk}

\section{Confidence Interval Eestimation of Overlap}

From Section 3, $\frac{\widehat{R}}{R}\sim F(2n_{1},2n_{2})$, then $\frac{\theta_{2}n_{2}}{\theta_{1}(n_{2}-1)}\widehat{R}^{*}\sim F(2n_{1},2n_{2})$. Let $L$ and $U$ be the lower and upper confidence limits respectively of $R$,  corresponding to the probability $1-\alpha$, i.e., $\mathbb{P}(L<R<U)=1-\alpha$.
Thus $L$ and $U$ can be determined by  solving for $R$ the equation 
$$ \mathbb{P}\left(F^{\alpha/2}_{(2n_{1},2n_{2})} <\frac{\theta_{2}}{\theta_{1}}\widehat{R}< F^{1-\alpha/2}_{(2n_{1},2n_{2})}\right)=1-\alpha $$
where $F^{\alpha/2}_{(2n_{1},2n_{2})}$  and $F^{1-\alpha/2}_{(2n_{1},2n_{2})}$
are the lower and the upper $\alpha/2$
quantile of the $F(2n_{1},2n_{2})$ distribution  respectively. Thus $$L=\frac{\widehat{R}}{F^{1-\alpha/2}_{(2n_{1},2n_{2})}} \quad and \quad U=\frac{\widehat{R}}{F^{\alpha/2}_{(2n_{1},2n_{2})}}$$

The lower $(L^{'})$ and upper $(U^{'})$ limits of OVLs can be obtained using appropriate transformation as $1-\alpha = Pr(L^{'} < OVL(R) < U^{'})$. Here $L^{'}=OVL(L)$ and $U^{'}=OVL(U)$. The confidence limits for OVLs are as follows:

\begin{table}[h]
\begin{center}
\begin{tabular}{ccccc}
\hline 
OVL & & lower limit $(L^{'})$ & & upper limit $(U^{'})$ \\ 
\hline 
$\Delta$ & & $1-L^{\frac{1}{1-L}}\vert 1-\frac{1}{L} \vert$ &  & $1-U^{\frac{1}{1-U}}\vert 1-\frac{1}{U} \vert$ \\ \\ 
\hline 
$\rho$ & & $\frac{2\sqrt{L}}{(L+1)}$ & & $\frac{2\sqrt{U}}{(U+1)}$ \\ \\
\hline 
$\lambda$ & & $\frac{4L}{(L+1)^{2}}$ & & $\frac{4U}{(U+1)^{2}}$ \\ \\
\hline 
$\Lambda$ & & $\frac{L}{L^{2}+L-1}$ & & $\frac{U}{U^{2}+U-1}$ \\ \\
\hline 
\end{tabular}
\end{center}
\end{table}

If $(L, U)\in (1 \infty)$, then the $L^{'}$ and $U^{'}$ interchange their role and the confidence interval for OVL becomes $(U^{'} ,L^{'})$ If 1 is enclosed in the interval $(L , U)$, then it asserts at $OVL=1$.

\section{Simulation Study}
A  Monte  Carlo  study  was  conducted  using to  evaluate  the  performance  of  approximations  to  bias  and  variance  of  four  overlap  coefficients.
From  each  population  $1000$  samples  of  $20,  50,  100$, $200$ and $500$  observations  were  generated. 
$\widehat{\rho}$, $\widehat{\lambda}$, $\widehat{\Delta}$ and $\widehat{\Lambda}$ were  computed  for each  pair  of  samples.  The  bias  and  variance  of  estimates  were  computed  using  actual OVLs  and  the  estimates.  The  bias  and  MSE  for  $R=0.2,  0.5,  0.8$  are  reported  in  Table  1. \\
The following conclusions are drawn based on these computations where only
the values of $R<1$ are considered. However, for the Overlap measures, the case $R<1$ is symmetric to the case $R>1$ the comments given below in terms of $R$ can also be interpreted in terms of $1/R$ for these OVL measures. 
\\
For sample sizes larger than 50, the bias is fairly close to zero. Weitzman's measure has less bias than others but Morisita's measure has the largest bias.
\\
The bias decreases as sample size increases, as expected and the MSE goes to zero for each OVLs.
$\Lambda$ tend to be more biased and the sampling distributions show larger variability. 
\\
It is clear that the  actual  OVLs  are  found  to  be  underestimated (Figure $3$) and for  very  small values of $R$ and small sample sizes, they are observe to be overestimated. The bias approaches $0$ very  fast.  For $n\geq 50$, the amount of bias is negligible and  fairly close to 0. Although $\widehat{\Lambda}$  has less bias than the other in case $R=0.2$ and has the largest bias for $R=0.8$; the bias of Delta approaches $0$ faster than the other three. The bias of $\widehat{\lambda}$ is the slowest in approaching $0$.  
\\
An important increase in standard deviations for small values of $R$ is observed for $\rho$ and $\lambda 
$. For $\Delta$ standard deviation increases as $R$ approaches $1$. But a remarkable increase in standard deviations for moderate values of $R$ in the $\Lambda$ case (Figure $4$). They  decrease  fast  as  $n$  increases, from $n=100$ the standard deviations are negligible. The  difference  between the  $MSE$  of $\rho$ and $\Lambda$ is  almost  nil  for  small  values  of  $R$,  but  the  difference  increases  as  $R$ becomes  large  with $\rho$ giving  lowest  $MSE$  values  and $\Lambda$ the  highest. \\
The estimates of MSE are plotted in Figure 5 for all four overlap coefficients.  As the sample size increases, the MSE reduces considerably.  
\begin{table}[h]
\begin{center}
{\small
\begin{tabular}{ccccccccccccc}
\hline 
  &  \multicolumn{3}{c}{$\widehat{\rho}$} & \multicolumn{3}{c}{$\widehat{\lambda}$}  & \multicolumn{3}{c}{$\widehat{\Delta}$}  & \multicolumn{3}{c}{$\widehat{\Lambda}$} \\ 
\hline 
$n$ & $Bias$  & $MSE$ & $Ratio$ & $Bias$ & $MSE$ & $Ratio$ & $Bias$ & $MSE$ & $Ratio$ & $Bias$ & $MSE$ & $Ratio$ \\ 
\hline
\\ 
c=0.2 & \multicolumn{3}{c}{$\rho =0.745$} & \multicolumn{3}{c}{$\lambda =0.556$} & \multicolumn{3}{c}{$\Delta =0.465$} & \multicolumn{3}{c}{$\Lambda =0.24$} \\ 
\hline 
20 & -0.029 & 0.007 & -0.36 & -0.030 & 0.016 & -0.25 & -0.0180 & 0.008 & -0.061 & 0.0060 & 0.0080 & 0.067 \\ 
50 & -0.011 & 0.003 & -0.22 & -0.012 & 0.006 & -0.15 & -0.0070 & 0.007 & -0.030 & 0.0020 & 0.0030 & 0.041 \\ 
100 & -0.055 & 0.001 & -0.15 & -0.056 & 0.003 & -0.11 & -0.0034 & 0.0015 & -0.017 & 0.0011 & 0.0015 & 0.029 \\ 
200 & -0.003 & 0.000$^{*}$ & -0.11 & -0.003 & 0.001 & -0.07 & -0.0020 & 0.027 & 0.010 & 0.000$^{*}$ & 0.000$^{*}$ & 0.020 \\
500 & -0.001 & 0.000$^{*}$ & -0.07 & -0.001 & 0.000$^{*}$ & -0.05 & 0.000$^{*}$ & 0.000$^{*}$ & -0.039 & 0.000$^{*}$ & 0.000$^{*}$ & 0.013 \\ 
\hline
\hline 
\\
c=0.5 & \multicolumn{3}{c}{$\rho =0.943$} & \multicolumn{3}{c}{$\lambda =0.889$} & \multicolumn{3}{c}{$\Delta =0.750$} & \multicolumn{3}{c}{$\Lambda =0.667$} \\ 
\hline 
20 & -0.036 & 0.0040 & -0.71 & -0.640 & 0.0140 & -0.66& -0.031 & 0.014 & -0.092 & 0.048 & 0.0500 & 0.22 \\ 
50 & -0.014 & 0.0010 & -0.44 & -0.024 & 0.0040 & -0.41 & -0.012 & 0.005 & -0.045 & 0.018 & 0.0190 & 0.013 \\ 
100 & -0.007 & 0.000$^{*}$ & -0.31 & -0.012 & 0.0020 & -0.28 & -0.006 & 0.0024 & -0.026 & 0.009 & 0.0090 & 0.095 \\ 
200 & -0.003 & 0.000$^{*}$ & -0.27 & -0.006 & 0.000$^{*}$ & -0.20 & -0.003 & 0.001 & -0.015 & 0.004 & 0.0045 & 0.067 \\ 
500 & -0.001 & 0.000$^{*}$ & -0.13 & -0.002 & 0.000$^{*}$ & -0.13 & -0.001 & 0.000$^{*}$ & -0.05 & -0.0018 & 0.0018 & -0.042 \\
\hline
\hline
\\ 
c=0.8 & \multicolumn{3}{c}{$\rho =0.994$} & \multicolumn{3}{c}{$\lambda =0.988$} & \multicolumn{3}{c}{$\Delta =0.918$} & \multicolumn{3}{c}{$\Lambda =0.952$} \\  
\hline 
20 & -0.032 & 0.001 & -0.87 & -0.063 & 0.005 & -0.87 & -0.037 & 0.016 & -0.3 & -0.20 & 0.061 & -0.84 \\ 
50 & -0.012 & 0.000$^{*}$ & -0.74 & -0.024 & 0.0011 & -0.73 & -0.014 & 0.006 & -0.19 & -0.079 & 0.013 & -0.69 \\ 
100 & -0.006 & 0.000$^{*}$ & -0.61 & -0.012 & 0.000$^{*}$ & -0.6 & -0.007 & 0.0027 & -0.133 & -0.039 & 0.005 & -0.56 \\ 
200 & -0.003 & 0.000$^{*}$ & -0.47 & -0.006 & 0.000$^{*}$ & -0.47 & -0.003 & 0.001 & -0.09 & -0.019 & 0.002 & -0.43 \\
500 & -0.001 & 0.000$^{*}$ & -0.32 & -0.002 & 0.000$^{*}$ & -0.32 & -0.001 & 0.000$^{*}$ & -0.06 & -0.008 & 0.000$^{*}$ & -0.28 \\ 
\hline 
\end{tabular} 
}
\end{center}
{\small $ ^{*}\vert value\vert < 0.001 $ $\quad ^{**}Ratio=Bias/\sigma $}
\end{table}

\begin{figure}[h]
\begin{center}
\label{Bias}
\includegraphics[scale=0.5]{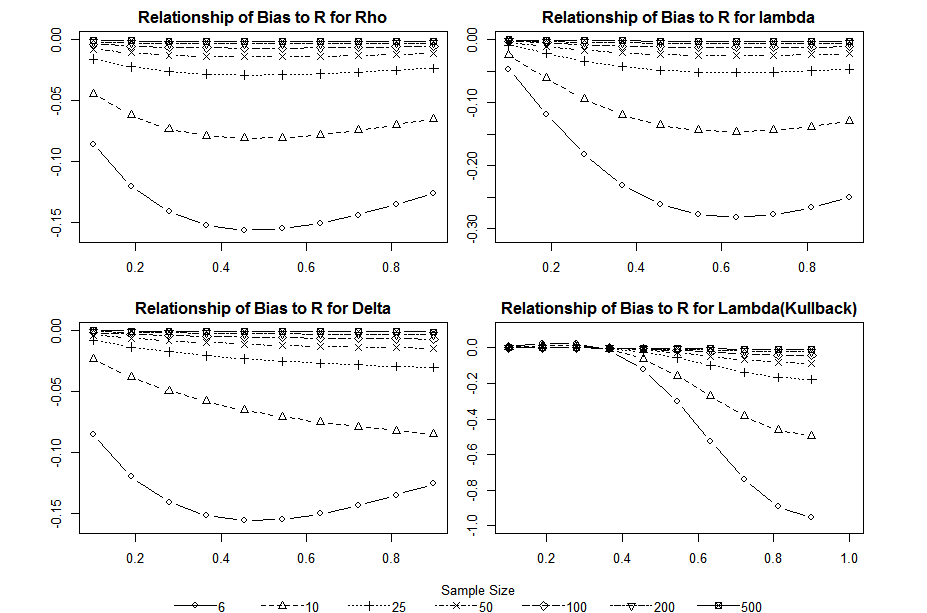} 
\end{center}
\caption{Relationship of Bias to $R$ for Overlap Coefficients.}
\end{figure}

\begin{figure}[h]
\begin{center}
\label{Std}
\includegraphics[scale=0.5]{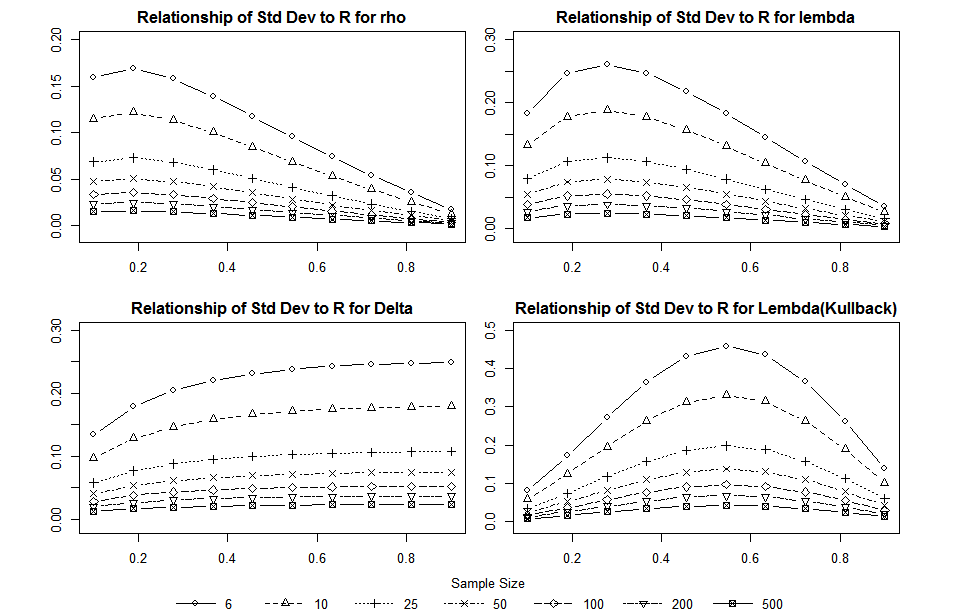} 
\end{center}
\caption{Relationship of Standard deviation to $R$ for Overlap Coefficients.}
\end{figure}

\begin{figure}[h]
\begin{center}
\label{Std}
\includegraphics[scale=0.5]{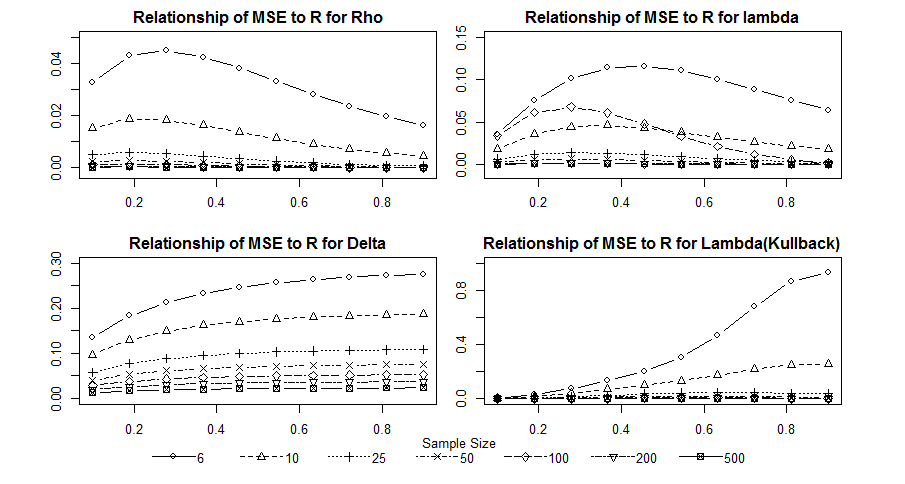} 
\end{center}
\caption{Relationship of MSE to $R$ for Overlap Coefficients.}
\end{figure}

\newpage

\section{Conclusion}
The  problem  of  estimation  of  four  commonly  used  measures  of  overlap  for  two exponential densities  with  heterogeneous  variances  is  considered  and  relations  between them  are  studied.  Overlap  coefficients  are  used  frequently  to  describe  the  degree  of interspecific  encounter  or  crowdedness  of  two  species  in  their  resource  utilization. 
\\
 Relations between  three commonly  used  measures  of  overlap with our measure of  overlap  are  studied  and  approximate  expressions  for  the  bias and  the  variance  of  the  estimates  are  presented.  The  invariance  property  and  a method  of  statistical  inference  of  these  coefficients  also  are  presented.  
Monte Carlo evaluations are used to study the bias and precision of the  proposed overlap measures.

\end{document}